\documentclass[sigconf]{acmart}

\usepackage{booktabs} 
\usepackage[flushleft]{threeparttable}
\usepackage{multirow}
\usepackage{adjustbox}
\usepackage{array}
\newcommand{\eg}{e.\,g., }
\newcommand{\ie}{i.\,e., }
\usepackage[belowskip=-10pt,aboveskip=3pt]{caption}

\copyrightyear{2018} 
\acmYear{2018} 
\setcopyright{othergov}
\acmConference[IVA '18]{IVA '18: International Conference on Intelligent Virtual Agents}{November 5--8, 2018}{Sydney, NSW, Australia}
\acmBooktitle{IVA '18: International Conference on Intelligent Virtual Agents (IVA '18), November 5--8, 2018, Sydney, NSW, Australia}
\acmPrice{15.00}
\acmDOI{10.1145/3267851.3267890}
\acmISBN{978-1-4503-6013-5/18/11}

\begin{document}
\title{Is Two Better than One?\\ Effects of Multiple Agents on User Persuasion}

\author{Reshmashree B. Kantharaju}
\affiliation{%
  \institution{ISIR, Sorbonne Universit\'e}
  \city{Paris} 
  \state{France} 
}
\email{bangalore_kantharaju@isir.upmc.fr}

\author{Dominic De Franco}
\affiliation{%
  \institution{School of Science and Engineering, University of Dundee}
  \city{Dundee} 
  \state{UK} 
}
\email{d.f.defranco@dundee.ac.uk}

\author{Alison Pease}
\affiliation{%
  \institution{School of Science and Engineering, University of Dundee}
  \city{Dundee} 
  \state{UK} 
}
\email{a.pease@dundee.ac.uk}

\author{Catherine Pelachaud}
\affiliation{%
  \institution{CNRS - ISIR, Sorbonne Universit\'e}
  \city{Paris} 
  \state{France} 
}
\email{catherine.pelachaud@upmc.fr}

 \renewcommand{\shortauthors}{R. B. Kantharaju et al.}
 \renewcommand{\shorttitle}{Is two better than one?}

\begin{abstract}
Virtual humans need to be persuasive in order to promote behaviour change in human users. While several studies have focused on understanding the numerous aspects that influence the degree of persuasion, most of them are limited to dyadic interactions. In this paper, we present an evaluation study focused on understanding the effects of multiple agents on user's persuasion. Along with \textit{gender} and \textit{status} (authoritative \& peer), we also look at type of \textit{focus} employed by the agent \ie \textit{user-directed} where the agent aims to persuade by addressing the user directly and \textit{vicarious} where the agent aims to persuade the user, who is an observer, indirectly by engaging another agent in the discussion. Participants were randomly assigned to one of the 12 conditions and presented with a persuasive message by one or several virtual agents. A questionnaire was used to measure perceived interpersonal attitude, credibility and persuasion. Results indicate that credibility positively affects persuasion. In general, multiple agent setting, irrespective of the focus, was more persuasive than single agent setting. Although, participants favored user-directed setting and reported it to be persuasive and had an increased level of trust in the agents, the actual change in persuasion score reflects that vicarious setting was the most effective in inducing behaviour change. In addition to this, the study also revealed that authoritative agents were the most persuasive.
\end{abstract}



\maketitle
\section{Introduction}

Intelligent virtual agents (IVA) have been incorporated into socio-technical systems which perform socially valuable functions, such as teaching~\cite{Swartout10AAG}, social coaching~\cite{Anderson13TTF}, and healthcare decision support~\cite{Devault14SKA}. As advances are made in their communicative abilities and social behaviour, the potential application and use of IVA's will shift away from a model of AI-as-tool to that of AI-as-assistant, AI-as-collaborator and AI-as-coach. Developing systems with an agenda, or {\em persuasive systems}, is now an active area of research~\cite{Benbasat10}, with one such class known as Behaviour Change Support Systems (BCSS). These are developed for the purpose of openly helping individuals or groups to change their behaviour: a BCSS is a ``a socio-technical information system with psychological and behavioural outcomes designed to form, alter or reinforce attitudes, behaviours or an act of complying without using coercion or deception''~\cite{Oinas-Kukkonen12}. Persuasion, in which attempts are made to create, modify, reinforce, or extinguish a user's beliefs, attitudes, intentions, motivations, or behaviour~\cite{Gass15persuasion}, is often integral to such systems. To develop persuasive BCSS's and measure their effectiveness, it is useful to draw on studies in argumentation and rhetorical strategies to understand the roles that group dynamics, argument types and characteristics of both persuader and persuadee play. In this paper, we aim to understand the effects of having multiple agents on user persuasion through an evaluation study by varying the agent's \textit{gender}, \textit{status} and \textit{focus} in both single and multiple agent settings. Results from this study will facilitate us in developing IVA's that will be able to handle conversations successfully and be effective in user behaviour changes. 

\section{Related work}
\subsection{Agent Characteristics and Persuasion}
Appearance plays an important role in influencing the user. Individuals are more influenced by agents who are similar to themselves with respect to appearance-related characteristics~\cite{Bailenson08SRI}; and in the context of a learning environment, an anthropomorphic agent with a human voice led to greater perceptions of agent credibility. For self-regulation and efficacy, gender played an important role. High self-regulation and self-efficacy was observed with those students who worked with mentor or motivator.

Real world gender stereotypes have been shown to be projected to computing environments~\cite{Carli01GAS} and has shown to be applicable to animated agents~\cite{Moreno02EAE}. Studies have shown that observers tend to be more influenced by an agent of same gender~\cite{Baylor04PAD, Guadagno07VHA}. Male instruction agent was preferred to work with over a female instruction agent~\cite{Kim07PAL}. Undergraduate women tended to choose to `learn about engineering' from agents who were male and attractive, but uncool conforming to the stereotypes. Students who worked with female agents showed higher self-efficacy beliefs than students who worked with the male agents because they perceived female agents as less intelligent~\cite{Baylor04PAD}. Studies have shown that female agents are subjected to more negative descriptions~\cite{Gulz10CGS, Silvervarg12TEO}. In~\cite{Kirkegaard14AGA}, a visually androgynous teachable agent, was used to study the influence of ascribed gender on perceived personality characteristics of the agent. Results revealed that the agent that was perceived as a girl received fewer positive words and more negative words than the same agent when it was perceived as a boy. In~\cite{Kramer16closing} virtual agents were used to enhance participants' performance, effort and motivation in mathematics. Results revealed that performance and effort were significantly enhanced when interacting with an agent of opposite gender.

The effect of gender on the persuasiveness (trust, credibility and engagement) using a robot and manipulating only the voice was done in~\cite{Siegel09PRT}. Cross gender interaction was observed for trust and credibility i.e., male participants trusted female robot more and vice-versa. Participants donated more often to a female robot than male robot. Further analysis revealed that male participants had a higher tendency to donate to a female robot while female participants had no preference. Men rated female robots as more credible and female rated male robots credible. The effects of gender and realism on persuasion was reported in~\cite{Zanbaka06CAV}. Participants were presented with a persuasive message regarding four topics, delivered by a male or female human, virtual human \ie an anthropomorphic agent, or virtual character \ie a 3D agent with ogre- or cat-like appearance. This study showed that participants found the virtual characters used in the study as persuasive as real humans. Visual realism of the speakers did not have an effect on the degree of persuasion. However, male participants were more persuaded by the female speakers than the male speakers, and female participants were more persuaded by the male speakers than the female speakers.

Based on the literature we can now say that agent gender, status is a significant attribute in interactions. Most of the existing studies mostly focus on effects of agent appearance and personality in learning environments and persuasion in dyads. 

\subsection{Argumentation and Rhetorical Strategies}\label{sec:Argumentation}
While much of the development in BCSS's has focused on subconscious methods of persuasion, recent theories such as the Transforming Sociotech Design model~\cite{tsd} have shifted the emphasis into enabling more permanent belief and behaviour change. As such, the dialogue and argumentation within it have an increasingly important role in facilitating these transformations. To assist with incorporating argumentation concepts into BCSS's a matrix has been developed~\cite{measuringpersuasiveness}. One of the most crucial of these concepts was introduced by Aristotle~\cite{barnes2014complete} -- the three fundamental modes of persuasion. \textit{Ethos} is primarily concerned with the character of the person making the argument and persuades by emphasizing the authority or credibility of the persuader. \textit{Pathos} relies on eliciting an emotional response from the persuadee and persuades with this appeal to emotion. \textit{Logos} is an appeal to reason and utilises evidence and sound argument to persuade.

The effects that the number of sources presenting a persuasive message has on attitude change has been studied from an information-processing view~\cite{Harkins81}. Harkins and Petty found that increasing the number of sources of a message increases thinking about the message content. In one experiment, subjects exposed to 3 compelling arguments presented by 3 different people were more persuaded than subjects exposed to the same number of arguments presented by just one person (and conversely, subjects exposed to 3 weak arguments presented by 3 sources were less persuaded than subjects exposed to the same number of arguments presented by just one person). The authority, credibility, importance or popularity of a speaker is well recognised in argumentation studies as affecting people's willingness to accept an argument. This is reflected in Aristotle's Ethos, as well as in common patterns of argumentation such as argument from authority, from popularity, and from expert opinion~\cite{Walton08}.

An important part of rhetoric is who the audience is. In many models of argument, the persuadee is considered to be an active participant in the argument, or a member of an audience in the case of a single orator. However, this is not always the case. The US televised Presidential Debates feature an argument-as-performance model, in which the goal of the debate is to persuade the audience rather than the opponent (in a recent study 29\% of people surveyed said the presidential debates were more helpful in helping them decide how to vote than anything else~\cite{holz2016presidential}). The legal system, in which lawyers argue -- apparently to persuade each other, but in reality to persuade a passive jury -- is another example of this model. These show the power of vicarious persuasion: the process where the aim is to persuade the audience rather than the person with whom a proponent is directly engaged in discussion. While understudied in argumentation, this power has been documented elsewhere, for instance in the domain of teaching (where vicarious experiences -- in which an individual observes another individual teach -- are one of the main sources that influence a preservice teacher's perceived self-efficacy~\cite{Bandura77}).

\subsection{Measuring the Effectiveness of a BCSS}\label{sec:persuasionQuestion}

The most influential approach for measuring the effectiveness of a BCSS in changing behaviour is the Persuasive Systems Design (PSD) Framework~\cite{oinas2009persuasive}. Based on the work of Fogg,~\cite{fogg2002persuasive} it suggests that the development of persuasive systems consists of understanding key issues, analysing the persuasion context and designing and analysing the system. To analyse a system, experts examine the BCSS against twenty eight design principles. These principles are split into four categories: primary task (personalising the system to the user, reducing effort on them and allowing them to self-monitor progress), dialogue (implementing computer-to-human dialogue to help users move towards their goal, including praise, rewards, reminders and suggestions), system credibility (aimed at making a system more persuasive through increasing its credibility through trustworthiness, expertise and authority) and social support (motivating users through social factors such as facilitation, cooperation and competition). 

Another method of measuring the persuasiveness of a BCSS after a period of user testing, is to issue the user with a questionnaire. An example of this is the Perceived Persuasiveness Questionnaire (PPQ)~\cite{lehto2012factors}. The PPQ was composed of twenty one questions relating to demographics, primary task support, dialogue support, perceived credibility, perceived persuasiveness, design aesthetics, unobtrusiveness and intention to continue the program. The three questions from the primary task section were utilised and adapted as were the three questions from the perceived persuasiveness section. A further three questions were taken from the perceived credibility section, although questions regarding professionalism and confidence were not relevant in the current experiment's scenario. 
A combination of these two approaches has generally been followed since De Jong et al showed consistent results between an expert PSD analysis, PPQ results and an analysis of user-test transcripts. This also matched with user log-data which showed consistent use of the BCSS studied~\cite{beerlage2014evaluation}. 

To measure how persuasive a system is, we must also take into account how persuadable the individual participants in our experiments are, since this factor alone could drastically vary the results of any research. Persuadability has been defined as the individuals' susceptibility to persuasive strategies and principles~\cite{kaptein2009can}. Based on Cialdini's~\cite{cialdini2007influence} six principles of influence; reciprocity, commitment and consistency, social proof, authority, liking and scarcity, Kaptein et al~\cite{kaptein2010individual} developed a 7-item persuadability instrument to determine participants persuadability score. Each item on the questionnaire is measured using a 7-point Likert scale, and average score is used to group participants into persuasion profiles ranging from low to high based on their scores. 

One must also consider the beliefs a user holds prior to a persuasive encounter and how the BCSS changed these beliefs. A useful way to measure belief change was suggested by Andrews et al~\cite{andrews2009measure} in which users rank their preferences before and after a persuasive interaction. A measure of persuasion is constructed and normalised from the difference between the two rankings. 

\section{Current Study}
The main objective of our experimental study is to understand the effects of using multiple agents in persuading users. We considered agent's characteristics (gender, status as displayed through different verbal and non-verbal behaviours) and focus (vicarious vs user-directed). We analysed the way these factors are perceived along the dimensions of credibility and interpersonal attitude. Through this study, we evaluate the effect of the following:

\begin{enumerate}
\item the {\bf gender} of the agents delivering the persuasive message
  ({\em male / female})
\item the {\bf number} of agents delivering the persuasive message
  ({\em 1 speaker delivering 6 arguments / 2 speakers delivering 3
    arguments each})
\item the {\bf status} of the agents delivering the persuasive message
  ({\em authoritative / peer})
\item the {\bf focus} of the agent delivering the persuasive message
  ({\em user-directed / vicarious})
\end{enumerate}

Through these we aim to answer the following research questions:
\begin{enumerate}
\item \textit{How significant is the effect of agent's gender and status on persuasion?}
\item \textit{Can use of multiple agents have a better positive outcome in persuading the users over having a single agent?}
\item \textit{Can vicarious persuasion be more effective than user-directed persuasion?}
\end{enumerate}

\section{Stimulus and Questionnaires}
In this section, we describe our experimental setup: the topic of persuasion, the agents used to present the persuasive dialogue and our design of the content. 

\subsection{Topic}
In order to avoid personal biases, the topic of discussion had to be as neutral as possible, while still being popular and broad. With this criteria in mind, we selected films as our discussion topic. To avoid any biases that may occur if our participants had previously watched any of the films, we created our own film descriptions. To ensure that the film descriptions themselves would not bias the outcome of our study we selected three different genres. A recent study on gender stereotypes in film~\cite{Wuhr17TOF} revealed the most popular and gender neutral genres. Once we had filtered for country and cultural biases, we were left with our 3 film genres: Comedy, Crime and History. The description of each of the 9 films followed a similar structure, while we kept the film titles and the language of the film descriptions as neutral as possible.

\subsection{Agent Appearance and Status}
\begin{figure}[h!]
\includegraphics[height=2in, width=2.2in]{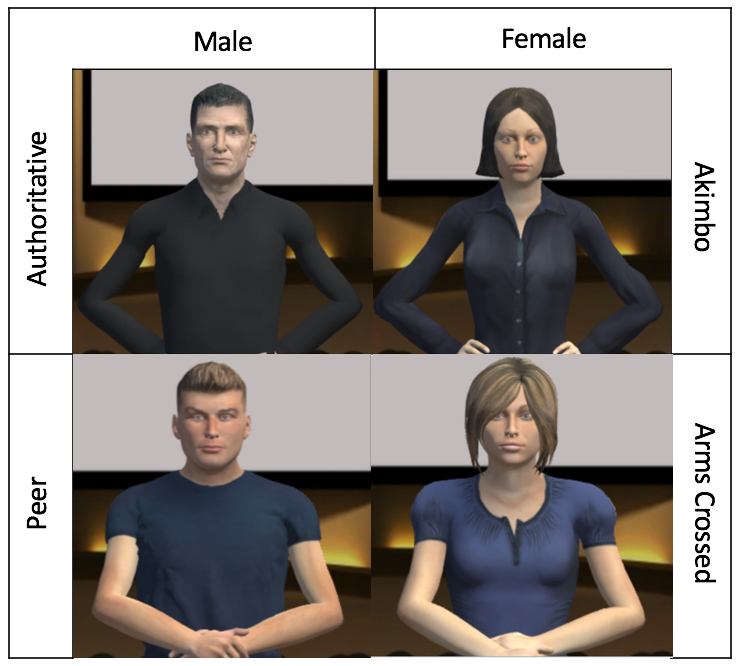}
\caption{Agent appearance}
\label{fig:appearance}
\end{figure}

Appearance, animations, affect and voice play an important role in defining the personality of the virtual agent~\cite{Baylor04PAD}. For this study, we designed four characters that differed in gender and status. We manipulated the visual appearance, non-verbal behaviours and linguistic style to fit the roles of an authoritative and a peer agent. The design of the appearance of the agents were largely based on literature that studied the effects of gender and status of virtual agents in motivating and learning environments~\cite{Baylor04PAD}. The authoritative agent was designed to fit the role of a film critique and to be perceived as an expert in films. Research shows that expertise in humans requires several years of deliberate practice in a domain~\cite{Ericsson93TRO}. Hence, we modeled the agent to appear aged in late-forties and dressed formally in a professional manner. The peer agent was designed to fit the role of a film-enthusiast who enjoys watching films and appeared as a student in the early-twenties and dressed casually. The appearances of the agents were designed using the Autodesk Character Generator software cf. Figure~\ref{fig:appearance}.

\subsection{Behaviours}
\label{sec:behaviours}

Body behaviour plays an important role in persuasive discourse and has been studied extensively~\cite{Kendon04GVA, Poggi08PAT,Poggi09GGA}. Although there are no particular gestures that could be categorised as persuasive, some gestures have a persuasive effect as they convey some of the information required in the persuasive structure of discourse~\cite{Poggi09GGA}. This information includes, importance, certainty, evaluation, sender's benevolence, sender's competence and emotion. Hence, in this study, along with verbal content, we also incorporated non-verbal behaviours that were expressed by the virtual agents. We chose to characterize agents' behaviours depending on their status only, either authoritative or peer. We did not differentiate in behaviours computation for the age or gender variables. As the agents don't differ much in age, we can assume their behaviour to not be affected by this variable. We did not want to fall into stereotypes and to assign typically gendered-behaviours. This is why we did not consider this variable either.

We define the peer agent as warm, friendly and the authoritative agent as competent and dominant. The chosen behaviours for the agents were based on literature on persuasion in humans~\cite{Poggi09GGA,Kendon04GVA}. In particular authoritative agents will display the \textit{Kendon's ring}~\cite{Kendon04GVA} and \textit{deictic finger point} (towards the interlocutor)~\cite{Poggi09GGA}. On the other hand, peer agents made use of more vague gestures such as \textit{beat open hand}. We also relied on studies that looked at the effects of behaviour on warmth and competence as well as attitude in virtual agents~\cite{Beatrice17ITR,Dermouche16sequence}. These studies informed us on the rest pose, facial expressions~\cite{Dermouche16sequence}, preferences to use either \textit{beat} gestures (rhythmic gestures not related to the semantic content of the speech) or \textit{ideational} gestures (more complex gestures related to the semantic content of the speech)~\cite{Beatrice17ITR}. Table~\ref{tab:behaviours} provides an overview of the variables we manipulated. We made use of the virtual agent platform Greta~\cite{Pecune14SFE} with Unity3D for generating the animations of the virtual agent and Cereproc TTS voice synthesizer to generate the audio for the virtual agents.
\newline

\begin{table}[h]
\centering
\caption{Overview of the distinctive characteristics for authoritative and peer agents.}
\label{tab:behaviours}
\begin{adjustbox}{width= 0.45\textwidth}
\begin{tabular}{|c|c|c|}
\hline
\textbf{Parameters} & \textbf{Authoritative} & \textbf{Peer} \\ \hline
Appearance & \begin{tabular}[c]{@{}c@{}}Aged\\ Formal clothing\end{tabular} & \begin{tabular}[c]{@{}c@{}}Young\\ Casual Clothing\end{tabular} \\ \hline
Facial Expressions & \begin{tabular}[c]{@{}c@{}}Frown\\ Corner lip down\end{tabular} & \begin{tabular}[c]{@{}c@{}}Eyebrow raise\\ Open smile\end{tabular} \\ \hline
Frequency of Smile & Low & High \\ \hline
Rest Position & Akimbo & Arms Crossed \\ \hline
Gestures & \begin{tabular}[c]{@{}c@{}}Ring\\ Deictic finger pointing \end{tabular} & \begin{tabular}[c]{@{}c@{}}Beat open hand \end{tabular} \\ \hline
\end{tabular}
\end{adjustbox}
\end{table}

\subsection{Dialogues}
Our dialogues utilised the three modes of persuasion (see Sec \ref{sec:Argumentation}). Each dialogue began with an introduction, before six arguments were made in front of the user. Two arguments were based on ethos, two on logos and two on pathos. Each dialogue then ended with a wrap-up section to conclude it. These dialogues were adapted for our variables giving us a dialogue for a single, peer agent, a single, authority agent, multiple agents using direct persuasion and multiple agents using vicarious persuasion. In the multiple agent scenarios using direct persuasion, the agents were both directing their arguments at the user with each using one ethos, pathos and logos argument. This scenario was split into three cases where we had two peer agents, two authority agents and one of each.

To distinguish between peer and authority approaches we incorporated ethos into the authority agent, used more formal language and vocabulary as well as utilised more credible sources. For the multi-agent dialogues using vicarious persuasion, one agent tried to persuade the other. This scenario was further split into four cases. In each case the persuading agent was the only one making an argument, while the other agent was listening and appearing as if they were being persuaded.
\newline

\begin{table}[h]
\centering
\caption{Examples of the dialogues used.}
\label{tab:Dialogues}
\small
\begin{tabular}{| m{4em} | m{6cm}|}
\hline
\textbf{Scenario} & \textbf{Dialogue} \\ \hline
Ethos Peer Comedy & I was listening to the radio the other night when the movie segment came on.
The movie critic Alex Garner said that [film name] was the funniest movie he's
ever seen. \\ \hline 
Ethos Authority Comedy & I noticed in the film section of this morning's paper that their film critic Alex
Garner gave [film name] the highest ever rating for a comedy. \\ \hline
Pathos Peer History & I was a wreck by the end of this film, [film name] had me high, then low, it
was like a roller-coaster! \\ \hline
Logos Peer Crime & If I were you I'd see this movie because it's full of suspense and drama as well
as being a good old fashioned crime movie. So you get a lot of bang for your
buck. \\ \hline
\end{tabular}
\end{table}

Argument structures both within and between the three genres were identical. Dialogues for each specific film differed only by film name, an incident, and a realisation, which were specific to each film. These were introduced systematically, with two references to the film name, one realisation and one incident for each genre and each argument type, in order to personalise the argument and avoid repetition for the participants. An example of the dialogues used can be found in Table~\ref{tab:Dialogues}.

\subsection{Evaluating Persuadability and Persuasiveness}\label{persuadability}
We used Kaptein's questionnaire on persuadability; however because it was written in the context of buying patterns and products, we adapted it slightly to involve characters discussing films. The first three questions were omitted from the persuadability measure in this experiment as they were not considered relevant to film choices and we wanted to avoid potentially overloading the participants with questions. We added a question concerning advice from trusted websites, as the ethos of the characters and the sources they recommended was a crucial part of this experiment. We evaluated both perceived persuadedness, as measured by a questionnaire based on the PPQ (described in Sec.~\ref{sec:persuasionQuestion}), and actual persuadedness, as measured by rating beliefs before and after interaction with the system. We adapted the ranking study described above (Sec.~\ref{sec:persuasionQuestion}) to use ratings\footnote{The actual ratings were between 1 to 5. We measured the change in this rating before and after the persuasive message and this change scales from -4 to +4.} as a number between -4 and 4, to give a more fine-tuned measure than the rankings used in~\cite{andrews2009measure}.

\section{Experiment}
\subsection{Design}
The experiment is based on 2 x 2 x 3 design. The variables include agent \textit{gender} (male vs. female), \textit{status} (authoritative vs. peer) and \textit{focus} (multiple agent user-directed vs. multiple agent vicarious vs. single agent). Since we are also studying the effects of gender, in multiple agent condition, a male agent and a female agent were present and only the status is altered. In vicarious persuasion, the status of both speaker and addressee agent will always be the same and only the gender is altered. Table~\ref{tab:condition} provides the overview of the twelve conditions\footnote{The verbal and non-verbal content remains the same across gender and conditions and varied only according to status.} used in the study. 
\newline

\begin{table}[h!]
\centering
\caption{Overview of the twelve randomly allocated experimental conditions; F: Female, M: Male, A: Authoritative, P: Peer.}
\label{tab:condition}
\begin{adjustbox}{width=0.45\textwidth}
\begin{tabular}{|c|c|c|c|}
\hline
\textbf{Focus} & \textbf{Composition} & \textbf{Condition} & \textbf{Participants} \\ \hline
\multirow{4}{*}{Single Agent} & F A & C1 & 18\\ \cline{2-4} 
 & M A & C2& 18\\ \cline{2-4} 
 & F P & C3& 18\\ \cline{2-4} 
 & M P & C4& 15\\ \hline 
\multirow{4}{*}{\begin{tabular}[c]{@{}c@{}}Multiple Agent\\ (vicarious)\end{tabular}} & F A persuades M A & C5& 18\\ \cline{2-4} 
 & M A persuades F A & C6& 17\\ \cline{2-4} 
 & F P persuades M P & C7& 18\\ \cline{2-4} 
 & M P persuades F P & C8& 16\\ \hline
 \multirow{4}{*}{\begin{tabular}[c]{@{}c@{}}Multiple Agent\\ (user-directed)\end{tabular}} & F A + M A & C9& 18\\ \cline{2-4} 
 & F P + M A & C10& 18\\ \cline{2-4} 
 & F P + M P & C11& 15\\ \cline{2-4} 
 & F A + M P & C12& 16\\ \hline
\end{tabular}
\end{adjustbox}
\end{table}

\subsection{Pre-Study Evaluation}
To assess the dialogues and in particular their mode of persuasion, we recruited two experts in argumentation and discourse analysis. The first expert (E1) had over 20 years experience in the communicative processes of argumentation, dialogue and persuasion, and the second one (E2) had 3 years experience as a post-graduate in the same area. We showed them our two mini-arguments for each mode and each of three film genres, i.e. 18 mini-arguments. These were presented in written form, in random order within the genre categories. We asked our experts to independently categorise the persuasion mode of each mini-argument as either ethos, pathos or logos, and  then gave them our model answers and asked them to check their categorisations against these and to let us know of any discrepancies and comments. Since we focused on persuasion mode we used only one status type - in this case all arguments were from our peer example. 

E1 warned us about attributing a single mode (ethos, pathos, logos) to dialogues, and highlighted the impact that the order in which different argument types are presented can have. E2 thought that some mini-arguments contained more than one mode of persuasion, and sent back a fully annotated set of the 18 arguments, with different parts of a mini-argument labelled ethos, pathos, logos. Based on the advice from these two experts, we simplified our mini-arguments to ensure that each argument was solely focused on one mode of persuasion. To avoid the impact that E1 warned us of, in the main experiment we presented these to participants in a randomised order.

To assess the agents along appearance and behaviours, we created a pre-study questionnaire and did a between-subject study with group ($n=12$), consisting of experts and na{\"\i}ve participants. We used static images of the four agents designed for the study and used a set of attributes to collect the first impression. The attributes included: above/below 30, student/professional, competent, friendly. For the authoritative agents, the attributes selected included above 30, professional, competent and for peer agents below 30, student, friendly. 

To assess agent's behaviour, we created two separate video clips using the same neutral~\footnote{Appearance is not modeled to fit any status} female agent displaying the non-verbal behaviours corresponding to both authoritative agent and peer agent ref. Sec~\ref{sec:behaviours}. A 5-point Likert scale was used to measure friendliness and competence. For the non-verbal behaviours, results showed a significant difference with agent friendliness ($p<0.05$), and even though the authoritative agent measured to be competent, it was not significant ($p>0.05$). With this study we ensured that the agents modeled fit to the status as well as their non-verbal behaviours.

\subsection{Method}
\subsubsection{Dependent Variables}
\label{sec:Dependent}
The dependent variables measured includes interpersonal attitude, credibility, persuasion ref. Sec~\ref{sec:persuasionQuestion}. We use a Likert-scale on a five-point scale ranging from 1 (Disagree) to 5 (Agree). To measure the \textit{interpersonal attitude}, we make use of the inter-personal circumplex proposed by Leary~\cite{leary04IDO}. The inter-personal circumplex is 2-dimensional, where affiliation (friendliness vs. hostility) is represented on one axis and status (dominance vs. submission) on the other axis. In total, the circumplex is divided into eight quadrants and we chose one adjective from each namely \textit{Assertive, Helpful, Warm, Un-authoritative, Timid, Distant, Arrogant, Forceful}. \textit{Credibility} was measured using 3 items developed by Kaptein et al~\cite{kaptein2010individual} on a 5-point Likert scale measuring perceived trustworthiness, reliability and expertise of the agent and \textit{perceived persuasiveness} measured using a 3 items questionnaire.

\subsubsection{Sample}
For this study we collected responses in two stages. Initially 282 participants were recruited from \textit{Crowdflower}. 156 responses were removed from the collected data due to inconsistencies and non-na{\"\i}vety as several participants did not adhere to the instructions and responded multiple times and we considered the responses to be not genuine. We also collected 79 responses by contacting respondents through mailing lists. In total, we had 209 participants where 55\% were male ($n=113$) and 45\% were female ($n=92$). 46\% of the participants were between the age range of 21-30 years, 22\% between 31-40, and 15\% between 41-50 and 14\% above 50 years old. The participants came from different cultural backgrounds with the three most prominent groups from, North America (37\%), Europe (27\%), and Asia (20\%).

\subsubsection{Procedure}
The participant began the study by filling in the demographics data \ie age, gender and education level followed by accepting the consent form. The study is divided into three main steps,(1) Pre-questionnaire, (2) Answering questionnaire after watching a video clip with persuasive dialogue (collected 3 times \ie one per given film genre), (3) Post-questionnaire. The pre-questionnaire is designed to measure the extent to which the participant is persuadable. Along with this the participant also provides information about overall openness and comfort towards technology and interest in films. A short introductory clip was designed using a virtual agent who presented the study. The age of the agent was in its 30s and its appearance was smart casual. This was done in order to familiarize the participants with the animations of the virtual agents to avoid collecting responses based on the first impression generated. 

The users are first presented with a short textual description of three films of a given genre and asked to rate the likeliness of watching the films respectively. Once the ratings are provided, the user is assigned randomly  to one of the 12 conditions specified above and presented with a persuasive video clip about the film. Since we want to measure the persuasion in user, we opted to show the clip corresponding to the film that received lowest rating by the user. The clip generally is 60s - 90s long, consisting of virtual characters presenting opinion and information about the film. The participants were again asked to rate the likeliness of watching the film again followed by questionnaire to measure attitude, perceived credibility, and perceived persuasiveness. This step is repeated again for the other two film genres. The condition does not differ between the genre of films and remains the same throughout the experiment. Finally, a post-questionnaire is used to measure persuasiveness, trust in the agents, overall satisfaction and intention to continue using the system. 

\subsection{Results}
In this section we present the results of the study and report on only the significant results from ANOVA (1-way and n-way).

\subsubsection{Perceived Attitude}
Attitude was measured using eight adjectives from Leary's interpersonal attitude circumplex ref. Sec.~\ref{sec:Dependent}. For each participant, we collected three responses, one after each film genre. Since the difference between the three responses was not statistically significant, we averaged the three responses to simplify the analysis. In terms of user's gender, male users reported agents to be more ``distant" in comparison to female users ($p=0.029$) as well as, perceived the agents to be arrogant ($p=0.030$) and forceful ($p=0.008$). There was no significant effect of the user's gender on the rest of the attributes. In terms of agent's status, authoritative agents were considered to be more forceful ($p=0.024$) compared to peer agents but they were also considered to be more helpful($p=0.0034$). There was no significant difference with regard to agent's gender on the perceived attitude. Agents were considered to be warmer ($p=0.041$) and helpful ($p=0.030$) in multiple agent setting over single agent setting.

\subsubsection{Credibility}
There was a statistically significant difference in the credibility of the agents: authoritative agents were considered more credible ($p=0.0006$) than peer agents (\ie trustworthy ($p=0.003$), reliable ($p=0.0007$) and shows expertise ($p=0.001$)). Male users reported agents to be more credible than female users ($p=0.025$). There was no significant effect of agent's gender on the perceived credibility. There was a statistically significant difference in credibility ($p=0.0003$) over the twelve conditions. We further performed ad-hoc test with Bonferroni correction. There was a statistically significant difference between following pairs: (C6 - C11 : $p=0.029$), (C9 - C11 : $p=0.002$), (C12 - C11 : $p=0.0007$). Figure~\ref{fig:credibility}, shows the mean perceived credibility value for each condition.

\subsubsection{Persuasion}
\label{sec:persuasion}

\begin{figure*}[h!]
\includegraphics[height=1.6in, width=\textwidth]{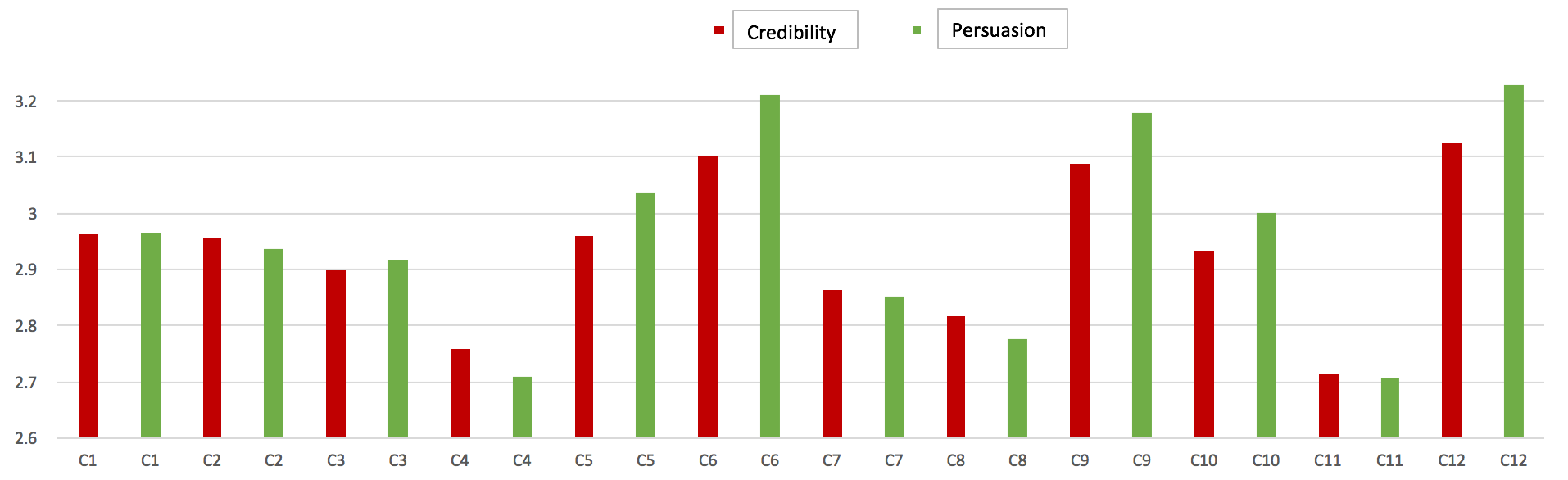}
\caption{Mean rating of \textit{perceived credibility} ($p=0.0003$)  C1-C4: Single agent ($m=2.894$); C5-C8: Vicarious ($m=2.936$); C9-C12: Multiple agent ($m=2.966$); and \textit{perceived persuasion} ($p=0.00002$) over 12 conditions. C1-C4: Single agent ($m=2.882$); C5-C8: Vicarious ($m=2.969$); C9-C12: Multiple agent ($m=3.028$); over 12 conditions.}
\label{fig:credibility}
\end{figure*}

Participants reported high persuasion with the multiple agent setting (C5 - C12) in comparison with single agent setting ($p=0.05$). Male user's reported high persuasion ($p=0.015$) compared to their female counterparts. However agent's gender did not have any significant effect on persuading the user. Authoritative agents were reported to be more persuasive than peer agents ($p=0.0004$). There was a statistically significant difference in persuasion ($p=0.00002$) over the twelve conditions. We further performed ad-hoc test with Bonferroni correction. There was a statistically significant difference between following pairs: (C4 - C9 : $p=0.048$), (C4 - C12 : $p=0.016$), (C6 - C11 : $p=0.015$), (C9 - C11 : $p=0.001$), (C12 - C11 : $p=0.0004$). Figure~\ref{fig:credibility}, shows the mean perceived persuasion value for each condition. Further, participants who scored to be '\textit{easily persuadable}' reported to be more persuaded ($p=0.0175$) than those who scored to be '\textit{not easily persuadable}'.
 
\section{Discussion}

The main focus of this research work was to understand whether having multiple agents was more effective than having single agent in persuasion task. We also wanted to understand the effects of gender and status on user-persuasion. In order to study this, we had three settings: single agent user-directed, multiple agents user-directed and multiple agents vicarious. 

The likeliness score of watching a film before and after the persuasive clip, is an indication that agents were successful in persuading the users to reconsider their decision about wanting to watch a film. 153 participants reconsidered their rating at least once and increased it. Participants who were grouped under `\textit{easily persuadable}' ($n=150$) reported significantly higher persuasion from both authoritative and peer agents and the change in scores indicate the same. Agent's status did not have any effect on participants grouped under `\textit{not easily persuadable}' ($n=55$), however, the vicarious setting was more effective in persuading them.

Authoritative agents were reported to be more credible regardless of the gender of the agent and participants reported higher level of trust in the information provided by them, indicating that the authoritative agents were perceived as competent~\cite{fiske2007universal}. This is in line with~\cite{Baylor05SIR}, where expert-like agents were perceived to be more credible. Additionally, they were also reported to be more persuasive than peer agents. In~\cite{holzwarth2006influence}, the expertise of the agent influenced the perceptions of credibility, and credibility mediated the influence of the agent's expertise on persuasion. 

While status played an important role, agent's gender did not have any significant effect. In previous studies~\cite{Baylor04PAD,Guadagno07VHA}, gender had a role in persuasion. However, in our study, gender was simply differentiated by the appearance of the agent and there was no other difference at the behaviour level or at the interaction level which can explain why there was no significant effect of gender. 

\begin{figure}[h!]
\includegraphics[height=2in, width=2.3in]{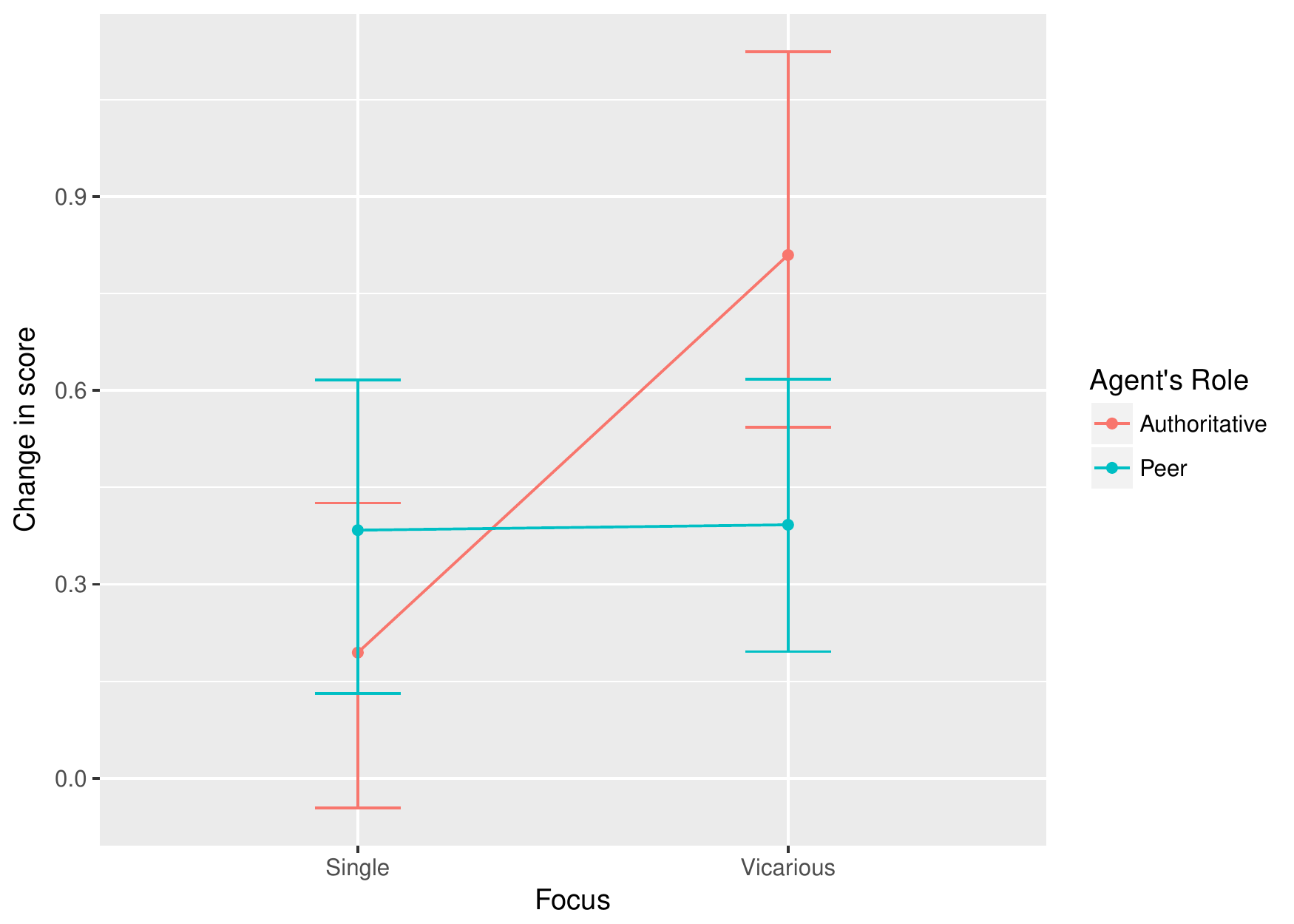}
\caption{Interaction of agent status$*$focus on change in rating ($p=0.035$). Authoritative agents were more persuasive in  multiple-agent setting (vicarious) than in single-agent setting.}
\label{fig:rating}
\end{figure}

The main finding of this evaluation study is that, a multiple agent setting was more effective than a single agent. The persuasiveness questionnaire revealed that participants reported being more influenced by the user-directed multiple agent setting. However, we measured the mean change in rating, for each condition and this revealed that vicarious setting was more effective in persuading the user to change their score than user-directed setting cf. Table~\ref{tab:meanvalues}. In particular, authoritative agents were more effective in vicarious setting than single agent setting cf. Figure~\ref{fig:rating}. Since the difference between the three settings was not statistically significant, we suggest that there is a strong tendency in the result that needs to be further verified with more participants.
\newline

\begin{table}[h]
\caption{Mean value of change in likeliness score of watching a film (before \& after the persuasive clip) and the self reported persuasiveness for the three conditions.}
\label{tab:meanvalues}
\begin{tabular}{|c|c|c|}
\hline
\textbf{Focus} & \textbf{Change in score} & \textbf{Self-report} \\ \hline
Single Agent & 0.285 & 2.882 \\ \hline
\begin{tabular}[c]{@{}c@{}}Multiple Agent\\ (Vicarious)\end{tabular} & \textbf{0.604} & 2.969 \\ \hline
\begin{tabular}[c]{@{}c@{}}Multiple Agent\\ (User Directed)\end{tabular} & 0.413 & \textbf{3.028} \\ \hline
\end{tabular}
\end{table}

Additionally, the agents in the multiple setting (user-directed) were considered to be more credible than a single agent and also users reported that they would consult the agents again and would recommend it to friends. This setting was also more helpful and users reported high satisfaction. 39\% of the users in single agent setting preferred to have multiple agents with different perspective while only 16\% preferred to have one agent condition. From the above results it is quite evident that multiple agent condition is indeed more effective, in particular, when vicarious persuasion is used.  

Our results on effects of settings are in line with human studies from social cognitive science. In \cite{meier2012language} it is argued that verbal persuasion by a single person is less efficient than vicarious experience on self-efficacy and behavioural change. Studies on interactive narrative systems report also that users are more influenced and engaged when experiencing vicarious social relationships and emotional responses than when experiencing events from their own direct environment \cite{slater2002entertainment}. Moreover, these studies underline how the effect of persuasion depends on the level of identification of the users with the interaction content. In our study, participants who reported being `\textit{easily persuadable}' did report high persuasion.

Further, the association between perceived credibility and perceived persuasion was observed using Spearman's rank correlation coefficient cf. Figure~\ref{fig:credibility}. We observe that perceived credibility positively affects the perceived persuasion ($r_s = 0.92$, $p<0.01$). This tendency has been studied in detail in~\cite{lehto2012factors,burgoon1990nonverbal,pornpitakpan2004persuasiveness}, where credibility is linked with persuasion. We can conclude that a credible agent can be effective for promoting behaviour change in a multiple agent setting.

\section{Limitations and Future Work}
Although the sample size for this study was lower, results indicate a significant relation between the type of persuasion used and agent attribute \ie status on user's persuasion. Future work will include collecting a larger sample of data and performing a detailed analysis focusing on the effects of status and persuasion type with multiple agents. Also, we aim to make the study more interactive, where the participant will be able to communicate with the agents and focus on studying the effectiveness of agents in various other domains \eg healthcare, education.

\begin{acks}
This project has received funding from the European Union's Horizon 2020 research and innovation programme under grant Agreement Number 769553. This result only reflects the authors' views and the EU is not responsible for any use that may be made of the information it contains. We are grateful to members of the Centre for Argument Technology at the University of Dundee for discussion of a previous draft, and in particular to Rory Duthie and Katarzyna Budzynska for detailed comments. We would also like to thank our anonymous reviewers for their thoughtful reviews.
\end{acks}

\bibliographystyle{ACM-Reference-Format}
\bibliography{sigproc.bib} 
\end{document}